# Doping Mn into $(Li_{1-x}Fe_x)OHFe_{1-y}Se$ superconducting crystals via ion-exchange and ion-release/introduction syntheses [*]

Huaxue Zhou(周花雪)[1,2,a], Shunli Ni(倪顺利)[2,3,a], Jie Yuan(袁洁)[2,3], Jun Li (李军)[4], Zhongpei Feng(冯中沛)[2,3], Xingyu Jiang (江星宇)[2,3], Yulong Huang(黄裕龙)[2,3], Shaobo Liu(刘少博)[2,3], Yiyuan Mao(毛义元)[2,3], Fang Zhou(周放)[2,3], Kui Jin(金魁)[2,3], Xiaoli Dong(董晓莉)[2,3,†] and Zhongxian Zhao(赵忠贤)[2,3]

[1]*College of Physics, Chongqing University, Chongqing 401331, China*
[2]*Beijing National Laboratory for Condensed Matter Physics, Institute of Physics, Chinese Academy of Science, Beijing 100190, China*
[3]*Key Laboratory for Vacuum Physics, University of Chinese Academy of Sciences, Beijing 100049, China*
[4]*Research Institute of Superconductor Electronics, Nanjing University, Nanjing 210093, China*

We report the success in introducing Mn into $(Li_{1-x}Fe_x)OHFe_{1-y}Se$ superconducting crystals by applying two different hydrothermal routes, ion exchange (1-Step) and ion release/introduction (2-Step). The micro-region x-ray diffraction and energy dispersive x-ray spectroscopy analyses indicate that the Mn has been doped into the lattice, and its content in the 1-Step fabricated sample is higher than that in the 2-Step one. Magnetic susceptibility and electric transport properties reveal that Mn doping influences little on the superconducting transition, regardless of 1-Step or 2-Step routes. By contrast, the characteristic temperature, $T^*$, where the negative Hall coefficient reaches its minimum, is significantly reduced by Mn doping. This implies that the reduction of the hole carriers contribution is obviously modified, and hence the hole band might have no direct relationship with the superconductivity in $(Li_{1-x}Fe_x)OHFe_{1-y}Se$ superconductors. Our present hydrothermal methods of ion exchange and ion release/introduction provide an efficient way for elements substitution/doping into $(Li_{1-x}Fe_x)OHFe_{1-y}Se$ superconductors, which will promote the in-depth investigations on the role of multiple electron and hole bands and their interplay with the high-temperature superconductivity in the FeSe-based superconductors.

## 1. Introduction

The newly discovered $(Li_{1-x}Fe_x)OHFe_{1-y}Se$ (FeSe-11111) superconductor has attracted extensive attention.[1-16] FeSe-11111 is composed by FeSe-tetrahedron layers and (Li/Fe)OH intercalation layers alternatively. Its superconducting transition temperature $T_c$, up to 43 K, is comparable with that of optimal $A_y$Fe$_{2-x}$Se$_2$ compounds ($A$=alkeli metal)[17-22], and without the troublesome $\sqrt{5} \times \sqrt{5}$ Fe vacancy ordered insulating phase which is always intergrown with the superconductivity in FeSe-tetrahedron layers of $A_y$Fe$_{2-x}$Se$_2$.[5] Besides, its electronic structure is quite similar to that of FeSe monolayer whose $T_c$ is above 65 K.[9] The available high quality big single crystals together with its highly two-dimensional electron character [6] ensure the FeSe-11111 system being propitious for unveiling the interplay of electronic anisotropy and high-$T_c$ superconductivity in the multiband FeSe-based superconductors.

[*]Project supported by National Natural Science Foundation of China (projects 11574370, 61501220), Frontier Program of the Chinese Academy of Sciences (Nos. QYZDY-SSW-SLH001, QYZDY-SSW-SLH008), the National Basic Research Program of China (projects 2013CB921700, 2016YFA0300301) and "Strategic Priority Research Program (B)" of the Chinese Academy of Sciences (No. XDB07020100).
[a]these authors contribute equally
[†]Corresponding author. E-mail: dong@iphy.ac.cn

For in-depth investigations on the multiple electron and hole bands, especially their interplay with high-$T_c$ superconductivity in FeSe-11111, the substitution of 3$d$ metals on the Fe-sites should be an efficient method to tune the carrier density and the Fermi surface topology. In fact, the doping effect of Mn was widely studied in the Fe-based family,[23-29] and, in some case, the Mn or other 3$d$ elements that doped into the blocking layer do not work as a pair-breaking center, but rather modify the carriers or magnetic ordering.[29] In the FeSe-11111 system, however, the effects of 3$d$ metal doping on the Fe-sites are still an open question.

In this work, we report that by applying ion exchange (1-Step) and ion release/introduction (2-Step) routes, manganese was successfully introduced, for the first time, into the lattice of (Li$_{1-x}$Fe$_x$)OHFe$_{1-y}$Se single crystals. The superconductivity was not strictly suppressed by Mn doping, regardless of 1-Step or 2-Step samples. The negative Hall coefficient indicated that electron carriers dominate the electrical transport over the whole range of measuring temperature. The characteristic temperature, $T^*$, where Hall coefficient reaches its minimum, is obviously reduced from about 130 K down to below 80 K, implying that the hole band might have no direct relationship with the superconductivity in (Li$_{1-x}$Fe$_x$)OHFe1-ySe superconductors.

## 2. Experiments

Figure 1 illustrates the syntheses process of Mn-doped (001)-oriented (Li$_{1-x}$Fe$_x$)OHFe$_{1-y}$Se via two different hydrothermal routes, that is, the ion exchange (1-Step) and ion release/introduction (2-Step). All the hydrothermal reactions were performed in stainless steel autoclaves of 25 ml capacity with Teflon liners. The synthesis progress of 1-Step was similar to that previous reported on the growth of (Li$_{0.84}$Fe$_{0.16}$)OHFe$_{0.98}$Se single crystal,[6] except that Mn powder was used as the starting material instead of Fe powder. The start materials of large K$_{0.8}$Fe$_{1.6}$Se$_2$ matrix crystal, 0.3 g Mn powder (Alfa Aesar, 99.95% purity), 4 g LiOHH$_2$O (Alfa Aesar, 99.996% purity) and 0.003 mol selenourea (Alfa Aesar, 99.97% purity) were mixed with 5 ml de-ionized water and loaded into the autoclave, and then the autoclave was tightly sealed and heated at 100 ºC for 72 hours. The nominal K$_{0.8}$Fe$_{1.6}$Se$_2$ matrix crystal was synthesized via the same method described in the supplemental material of our previous work.[6] For the 2-Step, the synthesis progress consists of two parts, namely, the ion-release and ion-introduction. In the ion-release part, we used nominal $K_{0.8}Fe_{1.6}Se_2$ crystal as matrix to get large superconducting FeSe crystal of (001) orientation as introduced elsewhere.[30] The second progress was followed with the routine of 1-Step but using FeSe crystal instead of $K_{0.8}Fe_{1.6}Se_2$. All the eventually obtained Mn-doped $(Li_{1-x}Fe_x)OHFe_{1-y}Se$ crystals were washed by de-ionized water for several times to remove soluble side products.

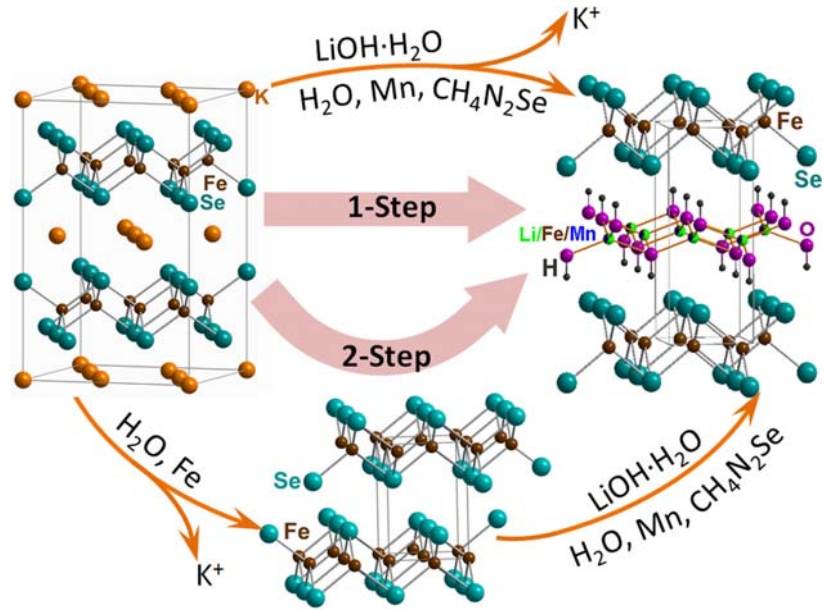

**Fig. 1.** Sketch of two different hydrothermal routes, ion exchange (1-Step) and ion release/introduction (2-Step).

X-ray diffraction (XRD) measurements of Mn-doped $(Li_{1-x}Fe_x)OHFe_{1-y}Se$ crystals was carried out at room temperature on a Rigaku SmartLab (9 kW) micro-region x-ray diffractometer with a ray source focusing 0.4 mm, with a $2\theta$ range of 5º–80º and a scanning step of 0.02º. In order to check the chemical composition both inductively coupled plasma atomic emission spectroscopy (ICP-AES) and energy dispersive x-ray spectroscopy (EDX) were carried out. The magnetic measurements were conducted on a Quantum Design MPMS-XL1 system with measuring field of 1 Oe to characterize the superconducting state. Both in-plane electrical resistivity and Hall resistivity data were collected on a Quantum Design PPMS-9 T.

## 3. Result and discuss

The energy dispersive x-ray spectra (EDX) are given in Fig. 2(a) and 2(b) The EDX analysis shows Mn has been successfully doped into the crystal. Quantitatively analysis is conducted by applying the inductively coupled plasma atomic emission spectroscopy (ICP-AES), which indicates that the atomic ratios of Mn:Fe:Se are (0.16:1.12:1) and (0.08:1.16:1) for the 1-Step and 2-Step samples, respectively.

Consequently, the content of Mn in 1-Step sample is considerably larger than in 2-Step.

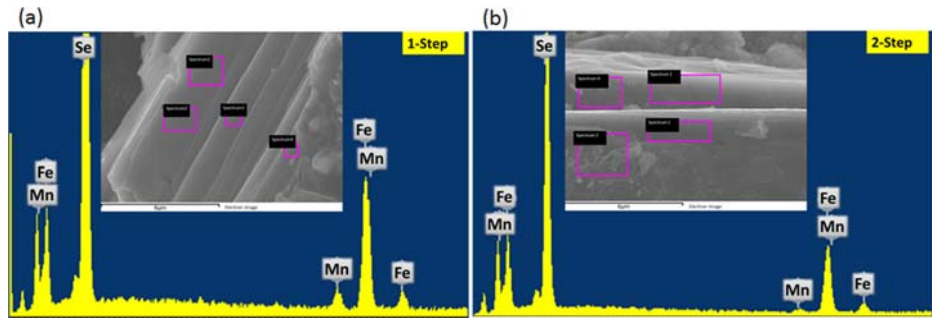

**Fig. 2.** (a, b) EDX analysis for samples obtained from 1-Step and 2-Step routes, respectively, showing the presence of elements Fe, Mn and Se. The insets show the corresponding SEM micrographs of the samples and regions where the data collected.

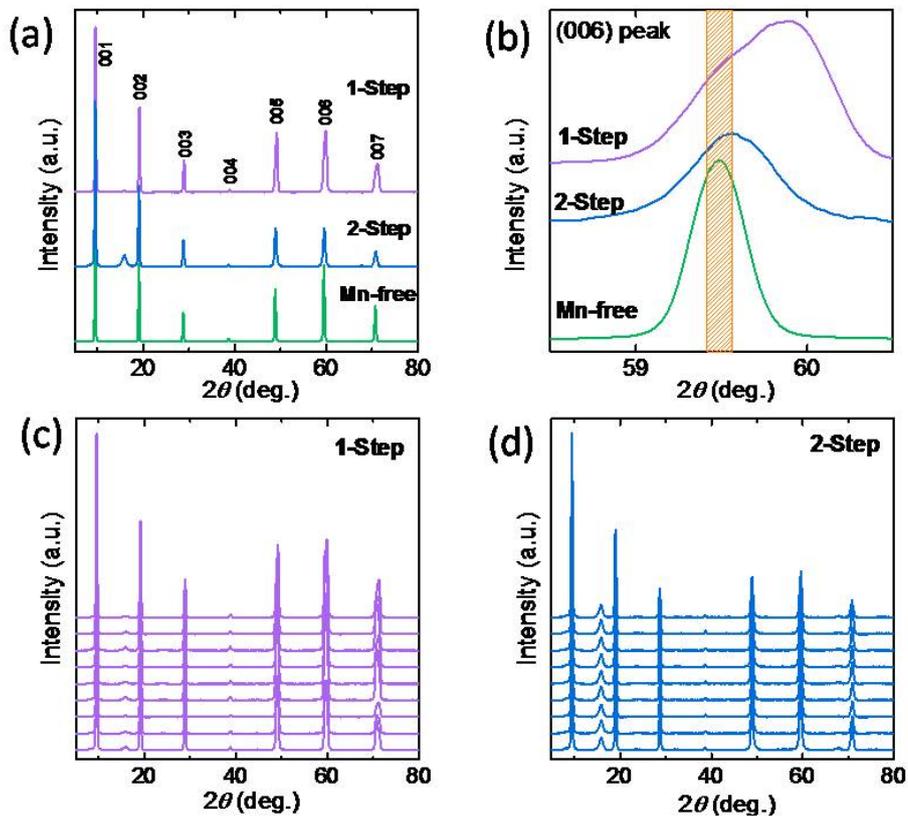

**Fig. 3.** (a) The XRD patterns of (00l) type for Mn-free $(Li_{0.84}Fe_{0.16})OHFe_{0.98}Se$ and Mn-doped $(Li_{1-x}Fe_x)OHFe_{1-y}Se$ crystals grown via ion exchange (1-Step) and ion release/introduction (2-Step). (b) Zoom-in (006) peaks, showing the position shift from left to right for 2-Step and more remarkable for 1-Step. (c) and (d) The XRD data collected uniformly in 9 different regions on the same sample for 1-Step and 2-Step sample, respectively.

Figure 3(a) shows the room temperature x-ray diffraction (XRD) patterns on 1-Step, 2-Step and Mn-free samples, demonstrating their crystal orientations along (00l) planes. The corresponding $c$-axis lattice parameters are 9.272(3), 9.306(3), and 9.322(1) Å. The enlarged view of (006) peak is shown in Fig. 3(b). With substitution of Mn, the (006) peaks show a pronounced rightward shift from the Mn-free $(Li_{0.84}Fe_{0.16})OHFe_{0.98}Se$. Considering the relatively smaller ion radius of the Mn than that of Fe, it is reasonable to contribute the lattice shrink to the substitution of Mn onto the Fe-sites, which is also in agreement with the ICP results. To check the homogeneity of Mn doping, we also collected XRD data at 9 different areas within a width of 0.4 mm for the 1-Step sample (Fig. 3(c)) and the 2-Step sample (Fig. 3(d)), respectively. All the diffraction patterns are well consistent, without obvious peak shift.

The temperature dependence of the static magnetic susceptibility for Mn-free $(Li_{0.84}Fe_{0.16})OHFe_{0.98}Se$, 1-Step and 2-Step crystals exhibits a sharp diamagnetic transition. The onset $T_c$'s for all crystals are almost identical (~ 40 K) with 100% diamagnetic shielding signal, as shown in Fig. 4(a). In Fig. 4(b), the temperature dependence of resistivity for the Mn-doped and Mn-free crystals shows that the zero resistance transition all occurs near 40 K as well. Obviously, the superconductivity of $(Li_{1-x}Fe_x)OHFe_{1-y}Se$ are not strongly sensitive to the Mn doping, being different from the previous results on other Fe-based superconductors.[23,24,26-29]

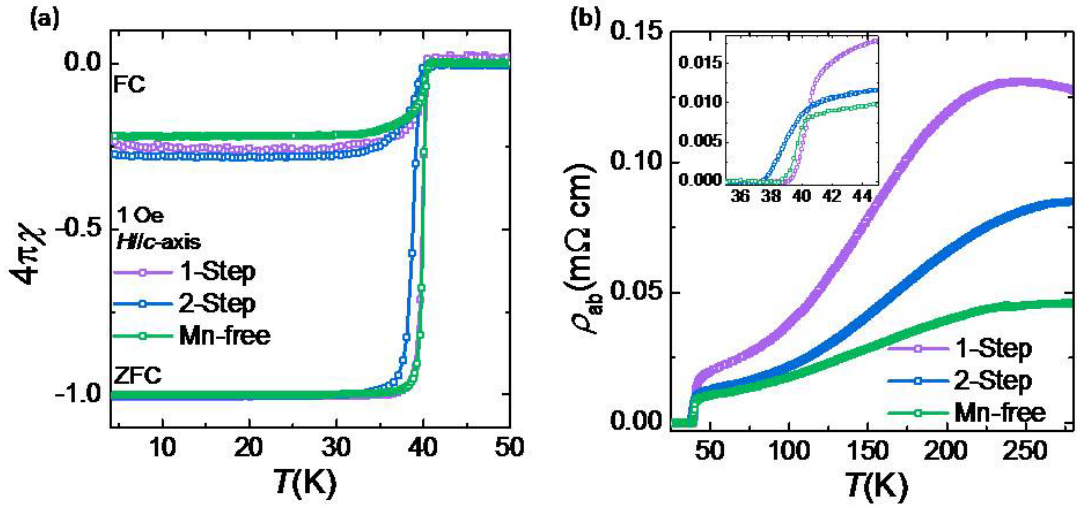

**Fig. 4.** (a) The magnetic susceptibilities of all crystals under zero-field cooling (ZFC) and field cooling (FC, $H$ along the $c$-axis) show that sharp diamagnetic transitions occur at around 40 K for all samples. (b) Temperature dependence of in-plane resistivity for all crystals. The inset is the zoom-in view around superconducting transition temperature.

Actually, as shown in Fig.5(a), the in-plane Hall resistivity ($\rho_{xy}$) for all crystals are also measured under the magnetic fields along the $c$-axis by sweeping the field up to 9 T at fixed temperatures. The extracted $\rho_{xy}$ is proportional to the magnetic field at all measuring temperatures. The negative Hall coefficient ($R_H$) indicated that electron carriers dominate the electrical transport over the whole temperature range for all crystals (Fig. 5(b)), while displays a minimum at a characteristic temperature, $T^*$. As previously reported,[6] above the $T^*$, the mobility of holes is reduced with decreasing temperature, leading to a decreasing Hall coefficient. Interestingly, the $T^*$ is dramatically reduced from about 130 K (Mn-free crystal) down to below 80 K (1-Step crystal, inset of Fig. 5(b)). This implies that the contribution of the hole carriers is obviously enhanced by Mn doping. Considering the fact that Mn doping shows no obvious influence on the superconductivity, we therefore suppose that the hole band has no direct relationship with the superconductivity in $(Li_{1-x}Fe_x)OHFe_{1-y}Se$ superconductors.

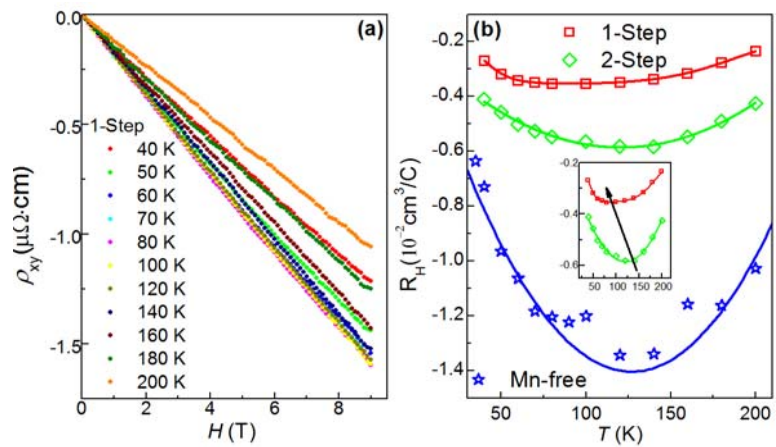

**Fig. 5.** (a,b) The Hall resistivity $\rho_{xy}$ and Hall coefficient $R_H$ as a function of temperature for Mn-free $(Li_{0.84}Fe_{0.16})OHFe_{0.98}Se$ and Mn-doped $(Li_{1-x}Fe_x)OHFe_{1-y}Se$ via 1-Step and 2-Step methods.

## 4. Conclusion

We have successfully doped Mn $(Li_{1-x}Fe_x)OHFe_{1-y}Se$ crystal via both ion release/introduction and ion exchange. The Mn was found to be doped into the lattice structure by both methods, while the concentration of Mn in the 1-Step fabricated sample is larger than that of 2-Step. Magnetic susceptibility and electric transport properties indicate that Mn doping influences weakly on the superconducting transition, while the $T^*$ is dramatically suppressed by Mn doping. It suggests that the hole band might have no direct relationship with the superconductivity in $(Li_{1-x}Fe_x)OHFe_{1-y}Se$ superconductors. Although we can hardly confirm the exactly doping position of Mn on the superconducting layer or the block layer, the present results suggested that hydrothermal methods of ion release/introduction and ion exchange can provide an effective way to explore bulk functional crystals desired for both basic research and applications, particularly for the universal mechanism of high-$T_c$ superconductivity.